\documentclass[aps,prd,twocolumn,superscriptaddress,reprint]{revtex4-2}

\usepackage{graphicx}
\usepackage{subfig}
\usepackage{ragged2e}
\usepackage{slashed}
\usepackage{lipsum}
\usepackage{xcolor}
\usepackage{verbatim}
\usepackage{physics}
\usepackage{hyperref}
\usepackage{cleveref}
\usepackage{appendix}
\usepackage{todonotes}

\definecolor{lime}{HTML}{A6CE39}
\DeclareRobustCommand{\orcidicon}{
	\begin{tikzpicture}
	\draw[lime, fill=lime] (0,0) 
	circle [radius=0.16] 
	node[white] {{\fontfamily{qag}\selectfont \tiny ID}};
	\draw[white, fill=white] (-0.0665,0.095) 
	circle [radius=0.005];
	\end{tikzpicture}
	\hspace{-2mm}
}

\foreach \x in {A, ..., Z}{\expandafter\xdef\csname orcid\x\endcsname{\noexpand\href{https://orcid.org/\csname orcidauthor\x\endcsname}
			{\noexpand\orcidicon}}
   }

\newcommand{\hl}[1]{}
\newcommand{\hnl}{\mathcal{N}}

\newcommand{\dmn}{d_{\mu\mathcal{N}}}

\begin{document}

\title{Constraints and Sensitivities for\\ Dipole-Portal Heavy Neutral Leptons from ND280 and its Upgrade}

\author{M-S. Liu\orcidA{}}
\email{msl63@cam.ac.uk}
\affiliation{Cavendish Laboratory, University of Cambridge, Cambridge, CB3 0HE, UK}
\author{N.W. Kamp\orcidB{}}
\email{nkamp@fas.harvard.edu}
\affiliation{Department of Physics and Laboratory for Particle Physics and Cosmology, Harvard University, Cambridge, MA 02138, USA}
\author{C.A. Arg\"{u}elles\orcidC{}}
\email{carguelles@fas.harvard.edu}
\affiliation{Department of Physics and Laboratory for Particle Physics and Cosmology, Harvard University, Cambridge, MA 02138, USA}

\begin{abstract}
    We report new constraints and sensitivities to heavy neutral leptons (HNLs) with transition magnetic moments, also known as dipole-portal HNLs.
    This is accomplished using data from the T2K ND280 near detector in addition to the projected three-year dataset of the upgraded ND280 detector.
    Dipole-portal HNLs have been extensively studied in the literature and offer a potential explanation for the $4.8\sigma$ MiniBooNE anomaly.
    To perform our analysis, we simulate HNL decays to $e^+e^-$ pairs in the gaseous time projection chambers of the ND280 detector and its upgrade.
    Recasting an ND280 search for mass-mixed HNLs, we find that ND280 data places world-leading constraints on dipole-portal HNLs in the 390-743\,{\rm MeV} mass range, disfavoring the region of parameter space favored by the MiniBooNE anomaly.
    The addition of three years of ND280 upgrade data will be able to disfavor the MiniBooNE solution at the $5 \sigma$ confidence level and extend the world-leading constraints to dipole-portal HNLs in the 148-860\,{\rm MeV} mass range.
    Our analysis suggests that ND280 data excludes dipole-portal HNLs as a solution to the MiniBooNE excess, motivating a dedicated search within the T2K collaboration and potentially highlighting the need for alternative explanations for the MiniBooNE anomaly. 
\end{abstract}

\maketitle

\section{Introduction}


Heavy neutral leptons (HNLs) are hypothetical right-handed counterparts to the left-handed neutrinos of the Standard Model.
They are a leading candidate for physics beyond the Standard Model (BSM), potentially explaining the origin of neutrino masses~\cite{Schechter:1980gr, Yanagida:1979as, Mohapatra:1979ia, Gell-Mann:1979vob, Minkowski:1977sc}, the nature of dark matter~\cite{Asaka:2005pn, Asaka:2005an, Bozorgnia:2024pwk}, and the matter-antimatter asymmetry in the Universe~\cite{Fukugita:1986hr, Davidson:2008bu, Abdullahi:2022jlv}.
In the minimal mass-mixing scenario, the HNL mass scale is unconstrained and typically determined by the Majorana mass scale of these additional right-handed states~\cite{Asaka:2005pn}.
The most effective experimental probe of HNLs changes with the mass scale; eV-scale HNLs are probed by neutrino oscillation experiments \cite{LSND:2001aii, MiniBooNE:2018esg, Ema:2023buz}, keV-scale HNLs by beta-decay searches \cite{KATRIN:2018oow} and X-ray spectra \cite{Hofmann:2019ihc}, MeV-scale HNLs by rare pion and kaon decay searches~\cite{Bryman:2019bjg, Bryman:2019ssi}, and GeV-to-TeV-scale HNLs by rare heavier meson decay searches and collider experiments \cite{Abada:2007ux, Cai:2017mow, Deppisch:2015qwa, Agrawal:2021dbo}.
Dipole-portal HNLs are an extension of mass-mixed HNLs characterized by a dimension-five effective transition magnetic moment coupling to photons and Standard Model neutrinos.
In the $\mathcal{O}(10-100)\;{\rm MeV}$ mass regime, HNLs with a nonzero dipole coupling to muon neutrinos have been explored as an explanation to the longstanding excess of electron-like events observed by the MiniBooNE experiment~\cite{MiniBooNE:2020pnu, Gninenko:2010pr, Gninenko:2011xa, Gninenko:2011hb, Gninenko:2012rw, Magill:2018jla}.
Specifically, the single-photon decay channel of these HNLs can mimic the single-electron signature of a $\nu_e$ charged-current interaction in MiniBooNE.
It has been shown that dipole-portal HNLs alone cannot explain the angular distribution of the MiniBooNE excess~\cite{Radionov:2013mca,Alvarez-Ruso:2021dna}.
However, Refs.~\cite{Kamp:2022bpt,Vergani:2021tgc} showed that dipole-portal HNLs in combination with a small contribution from eV-scale sterile neutrino oscillations can explain the angular distribution of the excess without suffering from significant tension in global sterile neutrino data.

The T2K near detector, ND280, provides an ideal environment to search for rare neutrino interactions due to its proximity to the high-intensity J-PARC neutrino source~\cite{Lux:2007zz, Lindner:2008yk, Ferrero:2009zz, Lindner:2008zz, T2K:2019bbb, Roth:2024mwe}.
It includes three low-density gaseous time projection chambers (TPCs), which can perform powerful low-background searches for dilepton decays of HNLs.
There are also fine-grained detectors (FGDs) comprised of segmented scintillator strips.
Dipole-portal HNLs can decay to $e^+e^-$ pairs through an off-shell photon, motivating a search for these HNLs in the ND280 TPC data.
This search can be improved in the upgraded ND280 detector, which will have five gaseous TPCs and a new super FGD comprised of scintillator cubes~\cite{T2K:2019bbb}.

In this letter, we use data from the 2019 T2K search for mass-mixed HNLs in ND280~\cite{T2K:2019jwa}, which includes $12.34\,(6.29)\times10^{20}$ protons-on-target (POT) in (anti-)neutrino mode, to place constraints on dipole-portal HNLs.
Additionally, we forecast the sensitivity of the ND280 upgrade, for which we assume $2\times10^{22}$ POT in total~\cite{Abe:2016tii}.
Ref.~\cite{Arguelles:2021dqn} excluded HNLs produced via mass-mixing in kaon decays in the ND280 decay pipe and detected via dipole-portal decays to single photons.
We use the same detection mechanism, but instead consider production via dipole-portal upscattering of neutrinos in the ND280 detector and upstream bedrock.
This was not possible in the prior study due to the lack of a detailed ND280 simulation.
However, in this study, we use \texttt{SIREN}~\cite{Schneider:2024eej} to simulate the interactions of HNLs with a dipole-portal coupling to muon neutrinos, $\dmn$, within a detailed model of the detector geometry.
We show that constraints from ND280 data disfavor the region of dipole-portal HNL parameter space preferred in the oscillation-plus-decay solution to the MiniBooNE anomaly introduced in Ref.~\cite{Kamp:2022bpt}.


The rest of this letter is organized in the following way.
In~\cref{sec:theory}, we describe the minimal HNL model and its dipole extension; in~\cref{sec:detector}, we introduce the ND280 detector and its upgrade; in~\cref{sec:simulation}, we describe the \texttt{SIREN}-based simulation of our signal events.
We present our main results in \cref{sec:results}.
Finally, in~\cref{sec:conclusion}, we offer some parting words of wisdom from this study. 
\section{DIPOLE PORTAL HEAVY NEUTRAL LEPTONS\label{sec:theory}} 

An HNL is a right-handed neutrino, $\hnl$, that transforms as a singlet under the Standard Model (SM) gauge group. 
The minimal HNL model interacts with the SM through mass-mixing. 
This model is typically considered in the context of the seesaw mechanism, which offers an explanation to the smallness of neutrino masses~\cite{Fernandez-Martinez:2022gsu}.
In the dipole extension, the HNL also couples to the SM lepton doublet, $L_\alpha = (l_\alpha, \nu_\alpha)$, and the electroweak gauge fields, $ B_{\mu \nu}$ and $W_{\mu \nu}^a$, through a dimension-six operator suppressed by a high-energy scale $\Lambda$. 
The relevant Lagrangian is given by
\begin{equation}
    \mathcal{L} \supset \frac{1}{\Lambda^2} \bar{L}_\alpha \tilde{H} \sigma^{\mu\nu} \mathcal{N} \left(C^\alpha_B B_{\mu\nu} + C^\alpha_W W^a_{\mu\nu} \sigma_a\right) + \text{h.c.},
\end{equation}
where $\alpha$ is the SM flavor index, $\sigma_a$ the Pauli matrices, $\tilde{H} = i\sigma_2 H^*$ the conjugated Higgs field, and $C_B^\alpha$ and $C_W^\alpha$ dimensionless coefficients. 
After electroweak symmetry breaking, this generates a dimension five \emph{transition magnetic moment} (TMM) term, also known as a dipole-portal operator,
\begin{equation}
    \mathcal{L} \supset d_{\alpha\mathcal{N}} \bar{\nu}_\alpha \sigma_{\mu\nu} F^{\mu\nu} \mathcal{N} + \text{h.c.},
\end{equation}
where \( d_{\alpha\mathcal{N}} = (v_h/\sqrt{2}) (C^\alpha_W C^\alpha_B + C^\alpha_W s_W)/\Lambda^2 \) and \( v_h \) is the Higgs vacuum expectation value.
This term couples the HNL to photons, providing a new phenomenological avenue for experimental observation. 
The HNL can similarly couple to other gauge bosons ($W$, $Z$), but their larger masses suppress these interactions.
We assume the dipole strength \( \vert d_{\alpha\mathcal{N}} \vert \gg G_F m_\mathcal{N}/(4\sqrt{3}\pi) \) so the massive-boson interactions are negligible.
In this study, we focus on a Dirac HNL with a dominant TMM coupling with the muon neutrinos, $\vert d_{e\mathcal{N}} \vert, \vert d_{\tau\mathcal{N}} \vert \ll \vert d_{\mu\mathcal{N}} \vert$.
There are thus two parameters relevant to this study: the HNL mass $m_\mathcal{N}$ and the TMM coupling $d_{\mu\mathcal{N}}$. 

In the dipole model, the HNL can produce two experimental signatures.
First, the electromagnetic decay of the HNL emits a photon through the process $\mathcal{N} \rightarrow \nu_\mu \gamma$, which is observable through pair-production off of a nucleus.
Second, virtual-photon-mediated decays can directly produce an $e^+e^-$ pair at the decay vertex.
For an HNL with $m_{\mathcal{N}} = 100\,{\rm MeV}$, the branching ratio for this process is $BR(\mathcal{N} \rightarrow \nu_\mu (\gamma^* \rightarrow e^+ e^-)) \sim 0.7\%$~\cite{Arguelles:2021dqn}.
The decay can be identified by the ND280 detector and its upgrade, both of which employ TPCs capable of identifying di-lepton HNL decays through two charged tracks, particularly $\mathcal{N} \rightarrow \nu_\mu (\gamma^* \rightarrow e^+ e^-)$~\cite{T2K:2019jwa, Arguelles:2021dqn}.
The FGDs, meanwhile, can detect on-shell photons from HNL decays, though they encounter significant backgrounds from neutral pion decays ($\pi^0 \rightarrow \gamma\gamma$). 
Although the upstream ND280 $\pi^0$ detector (P0D) can observe single-photon events, it is limited by a substantial background from low-energy photons. 
In contrast, the low-density gaseous argon TPCs provide a minimal background environment, making them optimal for HNL searches.
We leverage this idea to constrain dipole-portal HNLs via their decays to $e^+e^-$ pairs in the TPCs.
The main channel of HNL production in this analysis is Primakoff-like upscattering, $\nu_\mu A \to \mathcal{N} A$, off of nuclear targets in the bedrock and detector. 
\section{T2K and ND280 Upgrade} \label{sec:detector}

\begin{figure}
    \centering
    \subfloat{
        \includegraphics[width=0.45\textwidth]{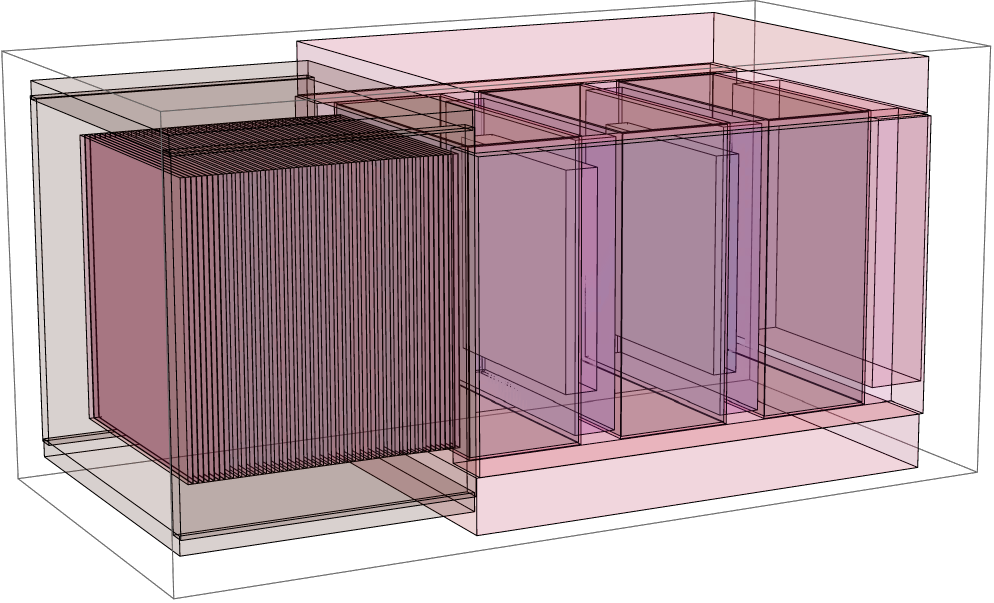}
        \label{fig:nd280}
    }\hfill
    \subfloat{
        \includegraphics[width=0.45\textwidth]{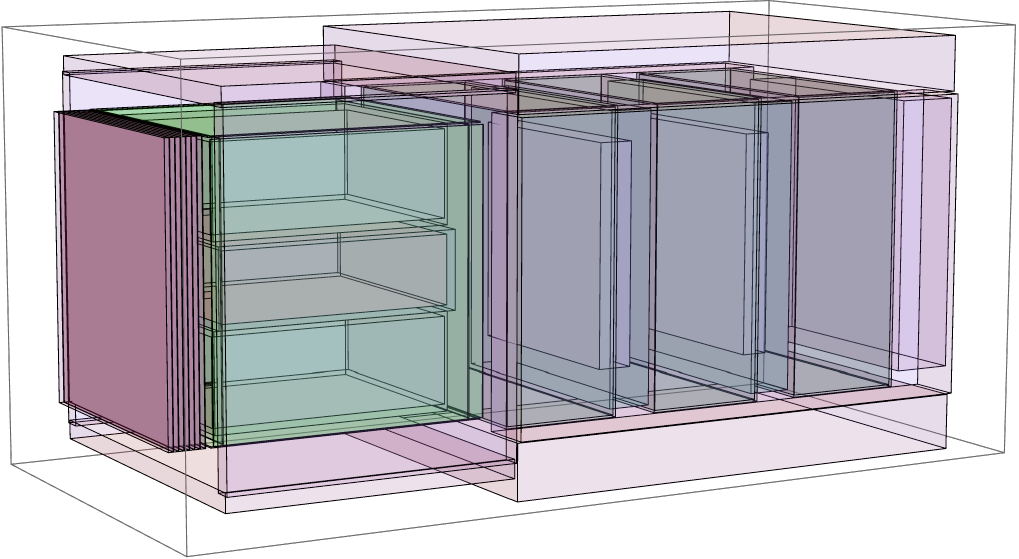}
        \label{fig:nd280+}
    }
    \caption{\justifying
    \textbf{\textit{Geometrical layout of the \texttt{SIREN} implementation of the nominal ND280 detector (top) and the upgraded detector (bottom).}}
    Both detectors share three downstream TPCs and two FGDs. In ND280, the upstream region houses the P0D. The upgrade replaces the P0D with the sFGD, sandwiched by two high-angle TPCs (HA-TPCs), keeping the upstream ECAL from the P0D detector.
    Each material is represented by a single color.}
      
    \label{fig:cad}
\end{figure}
The ND280 near detector of the T2K experiment is located 284.9 meters downstream from the J-PARC neutrino beam target in Tokai, Japan. 
This detector is essential for measuring the initial flux of electron and muon neutrinos ($\nu_e, \nu_\mu$), which enhances the sensitivity of both appearance and disappearance oscillation studies at the Super-Kamiokande far detector \cite{Blondel:2299599}. 
The T2K experiment initially took data from 2010 to 2021.
In 2023, the ND280 detector was upgraded for the T2K-II run, scheduled from 2023 to 2026. 
This upgrade increased the neutrino beam intensity by an order of magnitude, resulting in a projected $2\times10^{22}$ POT by the end of T2K-II~\cite{Abe:2016tii}. 
Additionally, the enhanced particle identification capability of the upgraded ND280 detector will help reduce systematic uncertainties in T2K flux characterization \cite{T2K:2019bbb, Dolan_2022}. 

\begin{figure*}
    \centering
    \includegraphics[width=\linewidth]{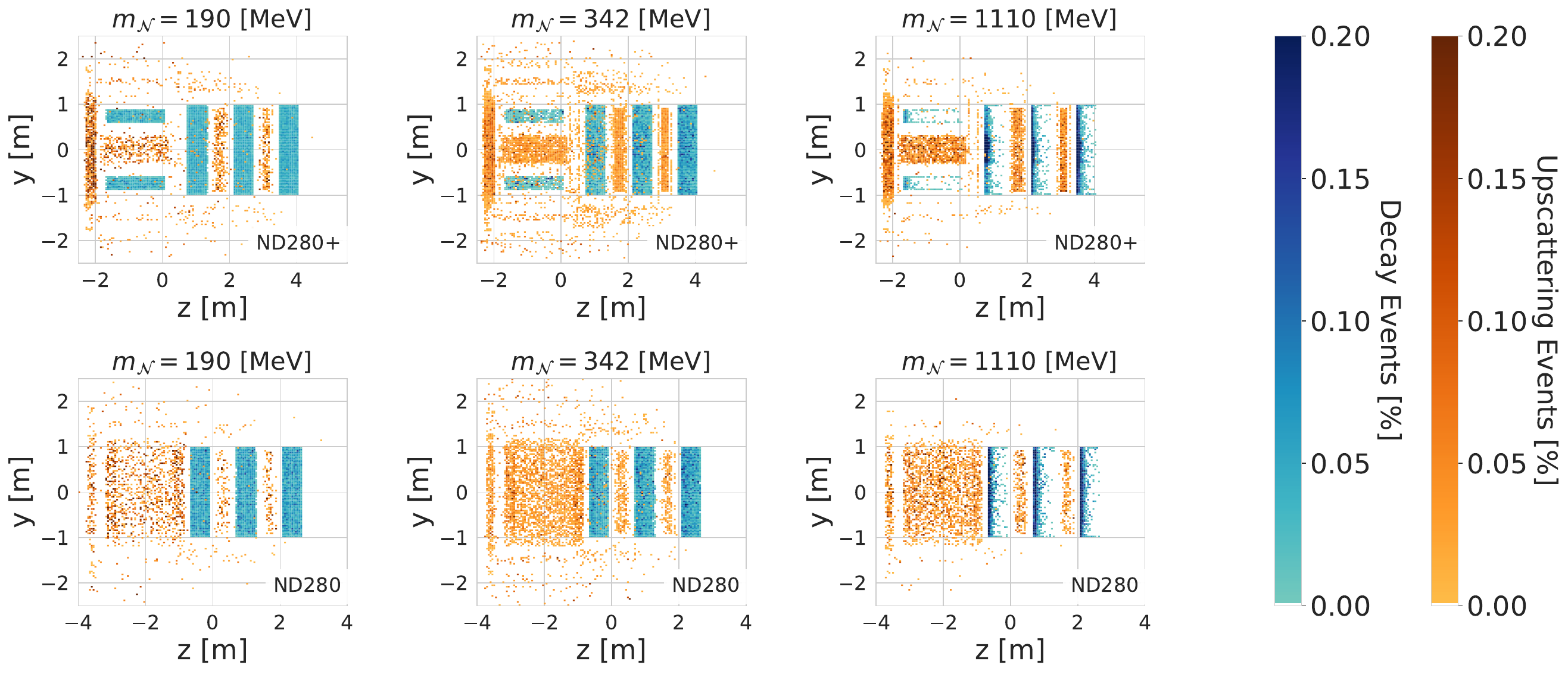}
    \caption{\justifying
    \textbf{\textit{Decay and up-scattering vertex of dipole-portal HNLs in ND280 and its upgrade for three different HNL masses.}}
    The two color scales are proportional to the weighted rates of HNLs whose decay products reach the fiducial volume.
    The top row is the ND280 upgrade, while the bottom is the ND280 detectors.
    All events are calculated assuming a dipole coupling of $d_{\mu\hnl} = 3.87 \times 10^{-7}$. 
    As the HNL mass increases from left to right, the effective decay length decreases, such that most of the signal comes from upscattering in the nearby detector sub-components.
    The T2K neutrino beam propagates along the z-axis.
    }
    \label{fig:decay}
\end{figure*}

We now review the components of the ND280 detector and its upgrade.
Both detectors share the same three downstream TPCs and two FGDs. 
The TPCs are filled with an argon-based gas mixture (95\% Ar, 3\% CF4, 2\% iC4H10).
The low density of the TPCs allows for precise tracking and momentum measurements of charged particles through their curvature in the 0.2 T magnetic field of the surrounding dipole magnet~\cite{T2KUK:2013wkh}. 
The TPCs also provide particle identification through $dE/dx$ measurements.
The higher-density FGDs serve as the primary neutrino interaction target. 
The upstream FGD1 consists of alternating layers of vertical and horizontal scintillator bars, while the downstream FGD2 has water layers interspersed between the scintillator layers. 
This design is crucial for extrapolating measurements of nuclear-target-dependent neutrino interactions from the near detector to the water-based far detector~\cite{T2KND280FGD:2012umz}.

The P0D occupies the upstream part of ND280. 
This detector sub-component consists of upstream and downstream electromagnetic calorimeter super-P0Dules with layered brass and scintillator bars, while the central super-P0Dules are layered with lead, scintillator bars, and water \cite{Ferrero:2009zz, Assylbekov:2011sh}.
This configuration is optimized for detecting neutral pions, a significant background source for electron neutrino appearance measurements.

The upgrade replaces the P0D with a SuperFGD sandwiched between two high-angle TPCs (HA-TPCs) \cite{T2K:2019bbb}. 
The SuperFGD consists of approximately 2 million optically isolated 1 cm$^3$ plastic scintillator cubes; each of which is read out by three wavelength-shifting fibers oriented along orthogonal axes~\cite{Kudenko:2022erj, Roth:2024mwe}. 
This design provides 3D tracking with improved spatial resolution, allowing for improved reconstruction of short particle tracks and better separation of final-state particles in neutrino interactions~\cite{Kudenko:2022erj}.
The HA-TPCs in the upgraded detector~\cite{Attie:2021yeh, Feltre:2023pcs, Attie:2023ttw} extend the angular acceptance for charged particle tracking, particularly for large-angle and backward-going tracks. 
This improvement is crucial for studying nuclear effects in neutrino interactions and for better characterizing the hadronic component of the final states.
The upgrade also includes six thin scintillator panels for Time-of-Flight (ToF) measurements~\cite{Korzenev:2021mny}, surrounding the SuperFGD and HA-TPCs.

For our study, we used the \texttt{SIREN} simulation package~\cite{Schneider:2024eej} to implement a detector model for ND280 and its upgrade.
These models are shown in \cref{fig:cad} and include the geometric layout and chemical composition of all the components described above.

\section{Simulating HNLs in T2K ND280 AND UPGRADE}
\label{sec:simulation}
Monte Carlo simulations of dipole-portal HNL interactions within the detectors were performed using the \texttt{SIREN} package \cite{Schneider:2024eej}, adapted from the \texttt{LeptonInjector} framework \cite{IceCube:2020tcq}.
\texttt{SIREN} models cascades of particle interactions with its \texttt{InteractionTree} object, allowing us to simulate both the production of HNLs via Primakoff upscattering and their decay to single photons.
We rely on the \texttt{DarkNews} interface of \texttt{SIREN}~\cite{Abdullahi:2022cdw} to compute the total and differential upscattering cross sections off of different nuclear targets, as well as the total and differential decay width of the HNL.
See Ref.~\cite{Kamp:2022bpt} and references therein for the full expressions of the cross section and decay width.
The detailed detector model interface of \texttt{SIREN} allows us to consider all of the different sub-components of ND280 and its upgrade outlined in \cref{sec:detector}.
This is essential for properly estimating the upscattering rate within each detector.

Each event is generated by first considering a muon (anti)neutrino traveling along the beamline with energy sampled according to the T2K $\nu_\mu$ ($\overline{\nu}_\mu$) flux reported in Refs.~\cite{T2K:2012bge, NA61SHINE:2018rhe}.
The upscattering location of this neutrino is sampled according to the interaction probability within each sub-component of the detector model along the line of sight.
We consider upscattering up to three HNL decay lengths in front of the TPCs.
Thus, for HNLs with decay lengths larger than the size of the detectors, we also consider upscattering in the upstream bedrock.
The kinematics of the outgoing HNL in the upscattering interaction are determined by sampling the differential cross section.
\texttt{SIREN} then simulates the decay of the HNL; the decay location is sampled using the lab-frame decay length of the HNL, and the kinematics of the outgoing photon and neutrino are determined by sampling the differential decay width.
To improve computational efficiency, \texttt{SIREN} requires HNLs to decay within the fiducial volume of the TPCs if possible.
We use \texttt{SIREN} to compute the physical weight of each upscattering-and-decay event.
From these weights, we can determine the total rate of single photons from HNL decays within ND280 and its upgrade.

\begin{figure}
    \centering
    \includegraphics[width=\linewidth]{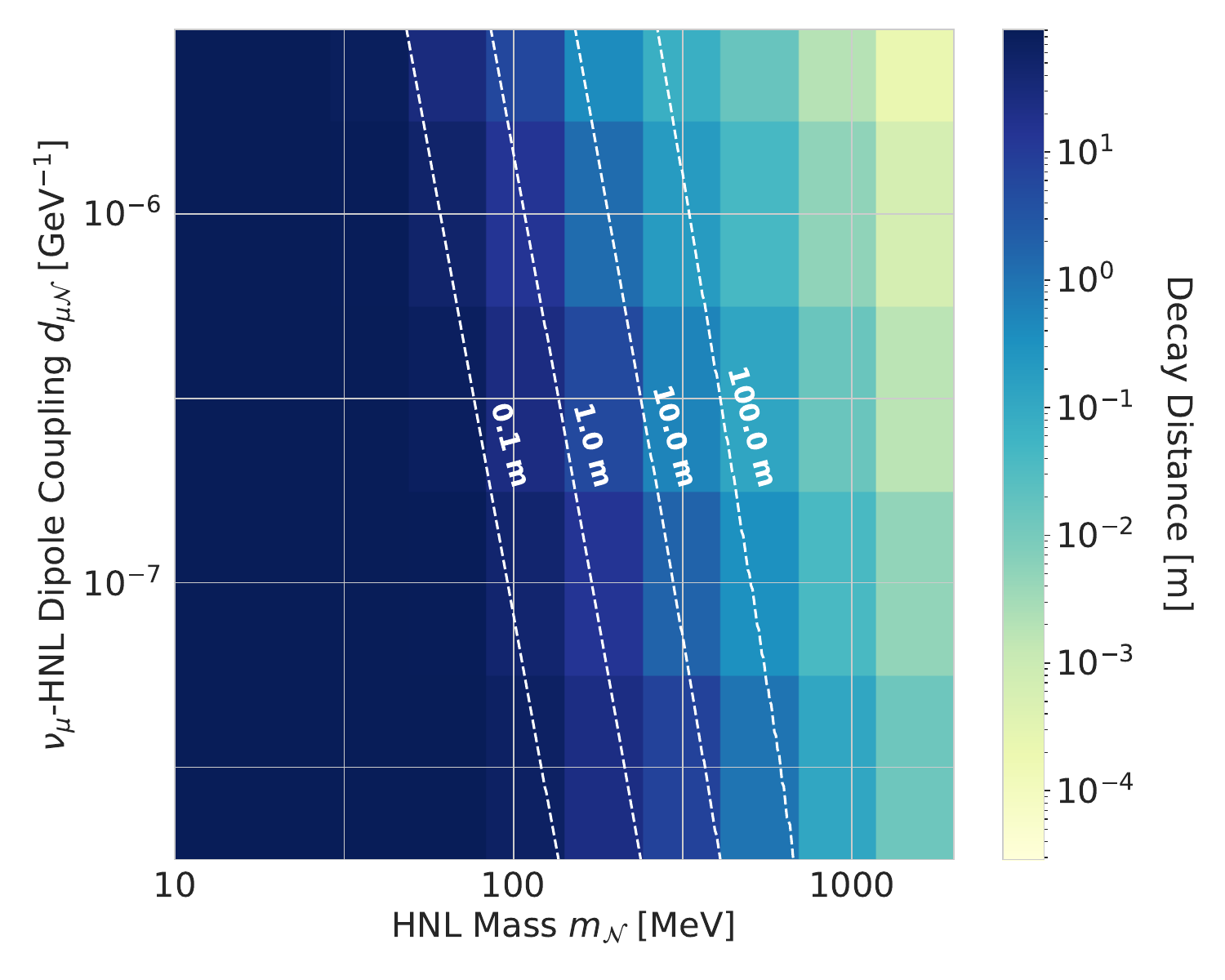}
    \caption{\justifying
    \textbf{\textit{The median decay distance of dipole-portal HNLs in ND280 at different points in the parameter space.}}
    The color axis reflects the prediction from the simulation described in \cref{sec:simulation}.
    The dashed lines are calculated analytically using the HNL decay width and the T2K flux.
    }
    \label{fig:distance}
\end{figure}

We inject $10^5$ neutrino events at a set of discrete points in $(d_{\mu \mathcal{N}},m_\mathcal{N})$ parameter space.
\Cref{fig:decay} shows the distribution of upscattering locations within ND280 and its upgrade for three of these points.
One can see that upscattering occurs almost entirely off of the high-density materials in each detector as opposed to the low-density gaseous TPCs.
Further, since the HNL decay length decreases as the mass increases ($\Gamma_{\mathcal{N} \to \nu \gamma} \propto m_\mathcal{N}^3$), one can see that most of the upscattering events for higher mass HNLs leading to single photons in the TPCs occur within the detector itself rather than in the (not shown) upstream bedrock.
In \cref{fig:distance} we show the median lab frame decay length of HNLs in our simulation across the parameter space.
There are two competing effects that will impact the sensitivity to dipole-portal HNLs.
As the decay length decreases, the sensitivity diminishes because HNLs decay before reaching the TPC. 
Conversely, as the decay length increases, the sensitivity diminishes because of the decreasing probability for HNLs to decay within the TPC.

The single photons produced in HNL decays will not pair-produce inside the gaseous TPCs; therefore, we must rely on dipole-portal decays directly to $e^+e^-$ pairs.
To do this, we down-weight each event in our simulation by the branching ratio ${\rm BR}(\hnl \rightarrow \nu_\mu (\gamma^* \rightarrow e^+ e^-))$~\cite{Arguelles:2021dqn}.
We sample the kinematics of the $e^+e^-$ pair according to eq.~A4 of Ref.~\cite{Arguelles:2021dqn}. 
The event selection applies the following criteria: (1) the opening angle $\Delta\Phi$ between the two charged tracks satisfies $\Delta\Phi < 90^\circ$, and (2) the polar angle $\theta$ of the incoming HNL is aligned with the beam direction ($\cos\theta > 0.992$).
We also consider an efficiency for tagging the $e^+e^-$ pair from dipole-portal HNL decays, which we assume to be the same as the efficiency for tagging $e^+e^-$ pairs from minimal HNL decay reported by T2K~\cite{T2K:2019jwa}.
It is important to note that this efficiency depends on the opening angle between the $e^+e^-$ pair, which is different for dipole-portal HNL decays and mass-mixed HNL decays.
We leave the detailed evaluation of the ND280 efficiency for tagging $e^+e^-$ pairs from dipole-portal HNLs for a future study.
Finally, we apply a fiducial volume cut within the TPC chambers, using the geometric parameters defined in Ref.~\cite{Lamoureux:2018owo}.


\section{DIPOLE PORTAL CONSTRAINTS} \label{sec:results}
\begin{figure}
    \centering
    \includegraphics[width=\linewidth]{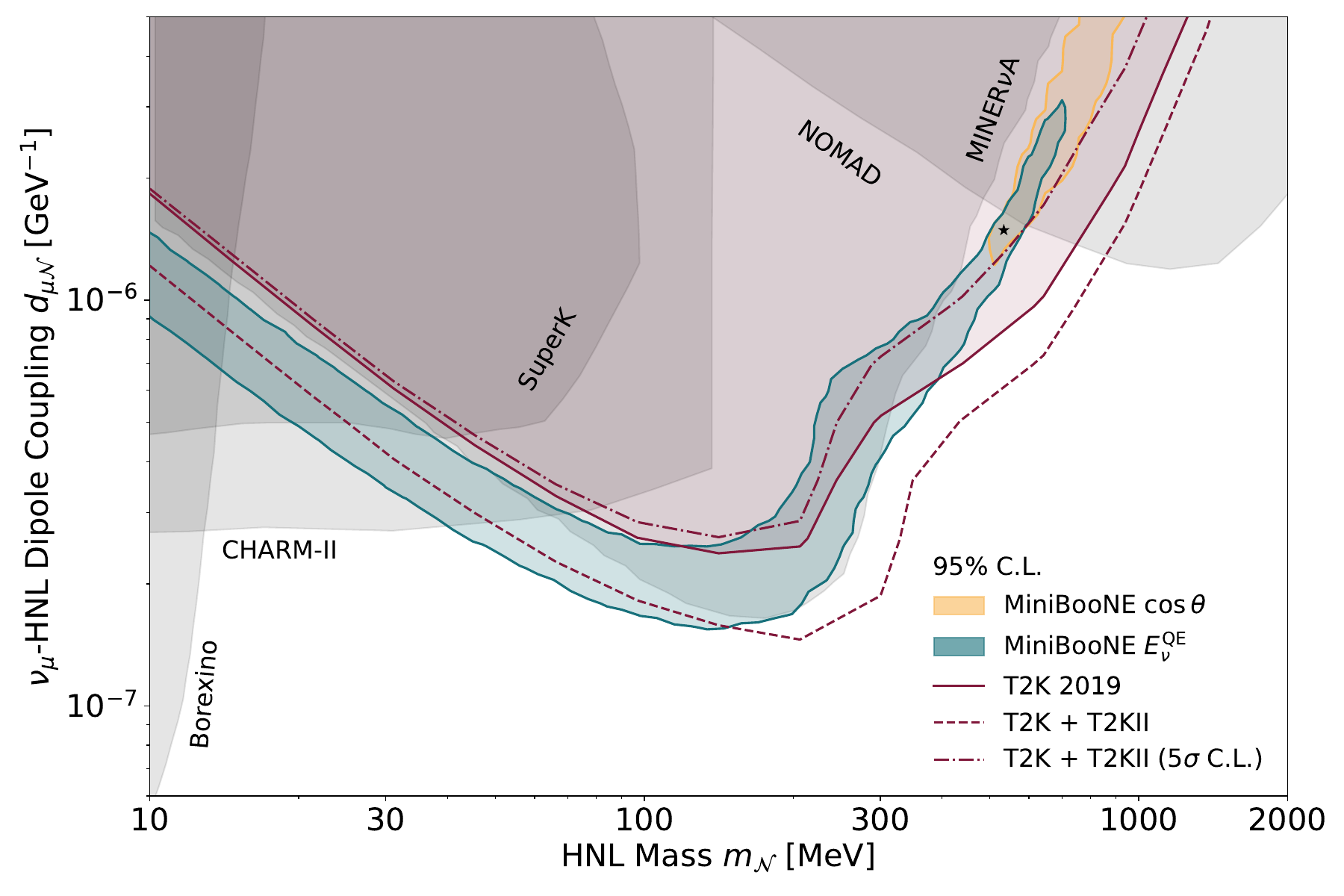}
    \caption{\justifying
    \textbf{\textit{Constraints and preferred regions at the 95\% C.L. in dipole-portal HNL parameter space.}}
    The results of this study are shown in red; the filled contour shows our constraints using the 2019 ND280 data~\cite{T2K:2019jwa}, while the dashed line shows the sensitivity of a combined analysis using the full ND280 dataset and three years of data from the upgraded ND280 detector.
    The preferred region to explain the energy and angular distributions of the MiniBooNE excess from Ref.~\cite{Kamp:2022bpt} are also shown.
    Our constraints disfavor this solution at the 95\% C.L.
    Assuming no excess of $e^+ e^-$ events emerges, the full T2K + T2K-II analysis will be able to rule out the MiniBooNE solution at the $5\sigma$ C.L.
    See the text for a description of the other constraints shown in the figures.
    }
    \label{fig:global}
\end{figure}

Prior to this work, several experiments have probed the region of dipole-portal HNL parameter space relevant to T2K.
Borexino sets limits on lower-mass HNLs ($m_\mathcal{N} \lesssim 10\,{\rm MeV}$) by detecting single-photon decays from neutrino upscattering within Earth, leveraging the large solar neutrino flux and the low energy threshold of the detector~\cite{Plestid:2020vqf}.
Similarly, Super-Kamiokande sets constraints on dipole-portal HNLs through the terrestrial upscattering of atmospheric neutrinos~\cite{Gustafson:2022rsz}.
CHARM-II \cite{Coloma:2017ppo} indirectly constrains dipole-portal HNLs up to $m_\mathcal{N} \sim 100\,{\rm MeV}$ through a measurement of the neutrino-electron elastic scattering cross section~\cite{Coloma:2017ppo}, and NOMAD directly constraints higher mass dipole-portal HNLs through a single photon search~\cite{Gninenko:1998nn}.
MINER$\nu$A sets strong constraints in the $m_\mathcal{N} \in [63\,{\rm MeV},486\,{\rm MeV}]$ regime through the high $dE/dx$ sidebands of their neutrino-electron elastic scattering analyses~\cite{Kamp:2022bpt}.
Finally, dipole-portal HNLs can potentially explain the long-standing excess of electron-like events observed by the MiniBooNE experiment~\cite{MiniBooNE:2020pnu, Gninenko:2010pr, Gninenko:2011xa, Gninenko:2011hb, Gninenko:2012rw, Magill:2018jla, Vergani:2021tgc}.
In this letter, we focus on the solution presented in Ref.~\cite{Kamp:2022bpt}, in which dipole-portal HNLs explain the bulk of the MiniBooNE excess alongside a smaller contribution from $\rm{eV}$-scale sterile neutrino oscillations.
Prior to this work, a portion of the preferred region from this solution at $m_\mathcal{N} \sim 500\,{\rm MeV}$ remained unconstrained at the 95\% confidence level (C.L.).
\Cref{fig:global} summarizes these constraints and preferred regions in dipole-portal HNL parameter space.

Using the simulation presented in \cref{sec:simulation}, we calculate constraints and sensitivity to dipole-portal HNLs from ND280 and its upgrade, respectively.
Following Ref.~\cite{T2K:2019jwa}, we rely on the non-observation of isolated $e^+e^-$ pairs in the gaseous TPCs of these detectors.
One can also search for single photons in the FGDs and SuperFGDs; however, their low event selection efficiency ($\sim1.9\%$) and poor signal-to-background ratio ($>1\%$) due to neutral pion decay backgrounds~\cite{T2K:2019odo} render them less effective at constraining the parameter space. 
We find that constraints from FGDs and SuperFGDs are over ten times weaker than those from the TPCs.
The gaseous TPCs, with their low single-shower backgrounds~\cite{T2K:2019odo, T2K:2019jwa}, provide the strongest constraints on dipole-portal HNLs.

Our 95\% C.L. constraints on dipole-portal HNLs from the 2019 T2K data~\cite{T2K:2019jwa} are shown in red in \cref{fig:global}.
These are derived using the Poisson likelihood to observe zero events given the simulation-derived signal prediction.
We ignore backgrounds in this calculation, which is conservative given the under-fluctuation observed in the ND280 data.
We find that the systematic uncertainties associated with the flux normalization and cross section are small ($\sim 5\%$) and do not significantly impact our results~\cite{T2K:2012bge}.
Thus, we leave them for a dedicated analysis by the experimental collaboration since the impact of flux uncertainties is likely smaller than that of a detailed assessment of the experimental efficiencies.
Two effects are apparent in \cref{fig:global}.
First, longer-lived HNLs in the lower mass regime are produced mostly through upscattering in the upstream bedrock.
The constraint from this production channel turns over once the decay length of the HNL is on the order of the fiducial volume size, at $m_\mathcal{N}\sim 200\,{\rm MeV}$.
Beyond this point, HNL production occurs mostly through upscattering in the detector itself.
This leads to the second ``bump'' in the observed constraint for $m_\mathcal{N} \gtrsim 300\,{\rm MeV}$.
Importantly, the constraint from this second region disfavors the dipole-portal HNL explanation for the MiniBooNE excess from Ref.~\cite{Kamp:2022bpt} at the $95\%$ C.L., as shown in figure \ref{fig:global}.

We also assess the capability of a full T2K + T2K-II analysis, using the full ND280 dataset as well as three years of data from the upgraded ND280 detector.
To do this, we compute the Asimov sensitivity within a Poisson likelihood, assuming that the analysis observes exactly the predicted background.
The search at the upgraded detector looks for $e^+e^-$ pairs in the original three TPCs as well as the two additional upstream HA-TPCs.
We conservatively assume that the background in the ND280 upgrade scales linearly with the POT and total TPC volume.
The dashed line in \cref{fig:global} shows the 95\% C.L. Asimov sensitivity of the T2K + T2K-II search.
Such an analysis is expected to exclude the MiniBooNE solution of Ref.~\cite{Kamp:2022bpt} at the $5\sigma$ C.L. after three years of T2K-II data.
It would also begin to probe previously unconstrained parameter space across a wide range of HNL masses.

\section{conclusion} \label{sec:conclusion}
In this letter, we set constraints on dipole-portal HNLs using data from a 2019 T2K search for $e^+e^-$ pairs from mass-mixed HNLs~\cite{T2K:2019jwa}.
We also evaluated the sensitivity of a combined T2K + T2K-II analysis using the full ND280 dataset as well as three years of data from the upgraded detector.
We used the \texttt{SIREN} Monte Carlo simulation to estimate the rate of $e^+ e^-$ pairs from dipole-portal HNL decay inside the gaseous TPCs of ND280 and its upgrade.
The detailed geometric treatment of our simulation enables the accurate calculation of the HNL production rate from Primakoff upscattering in the detector and upstream bedrock, an essential ingredient of the constraints presented here. 
Our main results, presented in \cref{fig:global}, show that the 2019 T2K data set world-leading constraints on the TMM coupling of HNLs with muon neutrinos, $d_{\mu \mathcal{N}}$, in the $m_\mathcal{N} \in [390\,{\rm MeV}, 743\,{\rm MeV}]$ region.
Importantly, they disfavor the dipole-portal HNL solution to the MiniBooNE excess presented in Ref.~\cite{Kamp:2022bpt} at the 95\% C.L.
Our sensitivity estimates further suggest that a full T2K + T2K-II analysis can set world-leading constraints on dipole-portal HNLs across a wider range of masses.

The results presented in this letter motivate an official search for dipole-portal HNLs within the T2K collaboration.
Just as the collaboration successfully leveraged the favorable setup of ND280 to set strong limits on mass-mixed HNLs~\cite{T2K:2019jwa}, so too can they use ND280 and its upgrade to reach untouched regions of dipole portal HNL parameter space.
Furthermore, at the level to which our simulation accurately reflects the ND280 detector, our constraints motivate alternative solutions to the MiniBooNE anomaly.
Studies like the one presented here are vital in paring down the expansive model space of MiniBooNE explanations~\cite{Acero:2022wqg}.
Thus, this letter represents an important step toward determining the nature of the anomaly.

\section{Acknowledgements}
We are grateful to Matheus Hostert for inspiring discussions and helpful input on the \texttt{DarkNews} parts of the simulation.
We are also grateful to the T2K collaboration, in particular Mathieu Lamoureux and Konstantin Gorshanov, for helpful discussions on the 2019 T2K results and intricacies of the ND280 detector.
Finally, we thank Alex Wen and Austin Schneider for input on the \texttt{SIREN} simulation.
Part of M-S.L.'s work was supported by the Cambridge Philosophical Society.
C.A.A. are supported by the Faculty of Arts and Sciences of Harvard University, the National Science Foundation (NSF), the NSF AI Institute for Artificial Intelligence and Fundamental Interactions, the Research Corporation for Science Advancement.
N.W.K. is supported by the National Science Foundation (NSF) CAREER Award 2239795 and the David and Lucile Packard Foundation. 

\bibliographystyle{apsrev4-1}
\bibliography{bib.bib}
\clearpage

\onecolumngrid
\appendix

\ifx \standalonesupplemental\undefined
\setcounter{page}{1}
\setcounter{figure}{0}
\setcounter{table}{0}
\setcounter{equation}{0}
\fi

\renewcommand{\thepage}{Supplemental Material-- S\arabic{page}}
\renewcommand{\figurename}{SUPPL. FIG.}
\renewcommand{\tablename}{SUPPL. TABLE}

\renewcommand{\theequation}{A\arabic{equation}}
\clearpage

\begin{center}
\textbf{\large Supplemental Material}
\end{center}

\begin{appendix}
    
\section{Methodology Validation}

\begin{figure}[ht]
    \centering
    \includegraphics[width=0.5\linewidth]{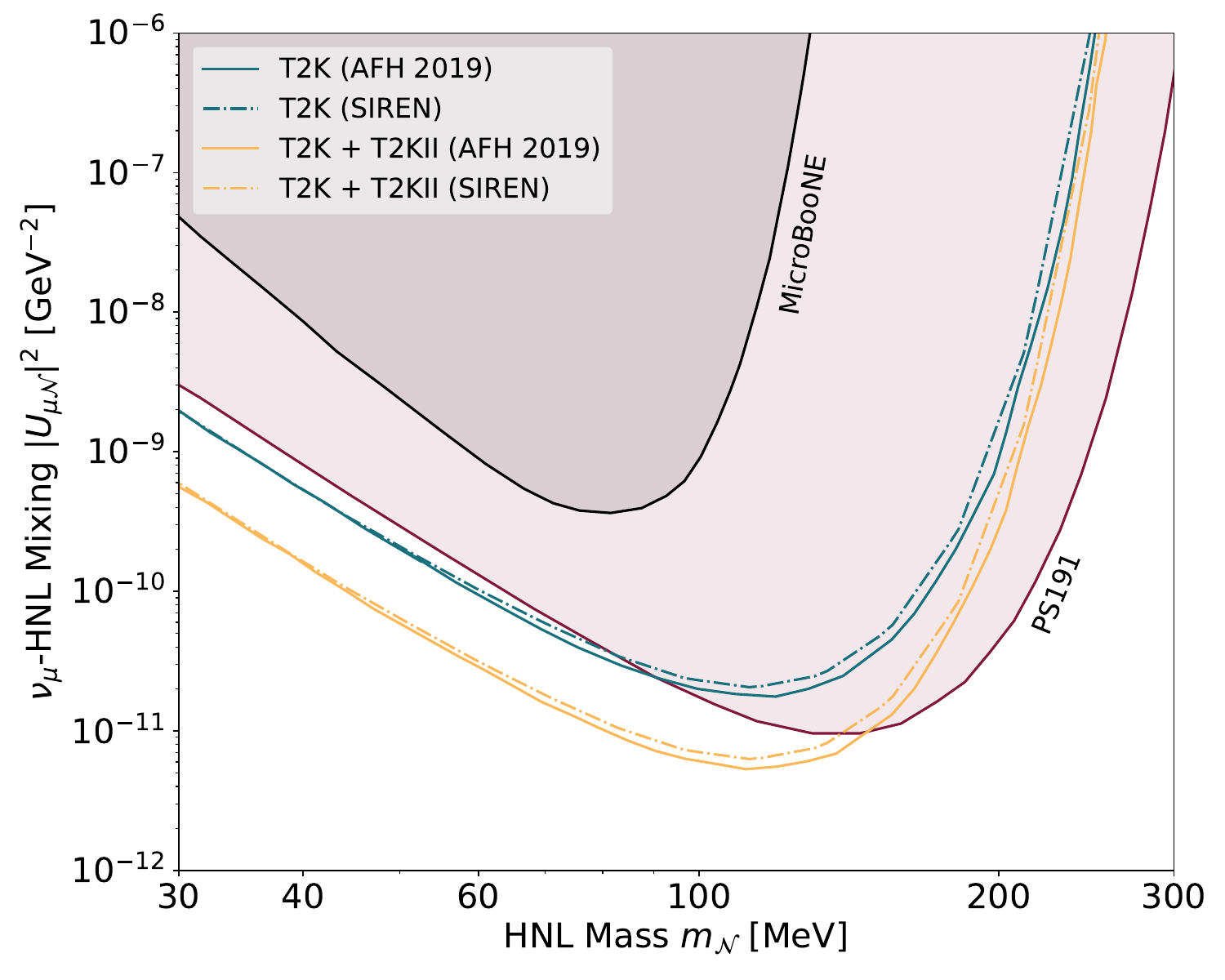}
    \caption{\justifying
    \textbf{\textit{Reproducing results from prior work.}}
    We reproduced the minimal mixing constraint $\abs{U_{\mu\hnl}}^2$ from Ref.~\cite{Arguelles:2021dqn} with $\mu_{tr}=10^{-6}$ using the di-electron decay search in the TPC and compared it to the \texttt{SIREN} fiducial volume cut for the same experiment, showing close agreement.}
    \label{fig:matheus}
\end{figure}
We validated our treatment of ND280 and its upgrade by reproducing the constraints from Ref.~\cite{Arguelles:2021dqn} on the minimal mixing parameter $\abs{U_{\mu N}}$ through TMM-mediated decays to $e^+ e^-$ pairs in the TPCs.
The dominant channel considered is the two-body decay $\hnl \rightarrow \gamma \nu_\mu$ via TMM. 
HNL decays to two leptons through a virtual photon or massive bosons are also included but are suppressed due to the additional vertex. 
The neutrino flux from kaon decays in T2K is provided in a T2K data release~\cite{T2K:2012bge}.
The flux of HNLs from kaon decays can therefore be calculated by scaling the spectrum down with the ratio~\cite{Shrock:1980vy}
\begin{equation*}
    \rho=\frac{\Gamma_{K\rightarrow \mu \hnl}}{\Gamma_{K\rightarrow \mu \nu_\mu}} = \abs{U_{\mu \hnl}}^2\,\frac{a+b-(a-b)^2}{a(1-a)^2}\, \lambda(1,a,b)^{1/2},
\end{equation*}
where $a=m_\mu/m_K,\,b=m_N/m_K$ and $\lambda$ is the K\"{a}ll\'{e}n function.
The resulting HNL flux is injected into our \texttt{SIREN}-based description of ND280 and its upgrade.
We then compute the number of decays that occur within the fiducial volumes summarized in \cref{tab:fid}.
\Cref{fig:matheus} compares our results to those obtained in Ref.~\cite{Arguelles:2021dqn}, showing good agreement for both the T2K 2019 constraints and the T2K + T2K-II sensitivity. 
Constraints are calculated at a 90\% confidence level, assuming zero background in the gaseous TPCs. 
These constraints are limited by the HNL survival probability from the beam target at higher masses and by decay probability within the TPC fiducial volume at lower masses.
\begin{table}[hb]
    \centering
    \begin{tabular}{|lccc|}
    \hline
    & x [m] & y [m] & z[m]\\
    \hline
        HA-TPC & 1.70&1.96&0.56  \\
       TPC &1.59& 0.27& 1.71 \\
       \hline
    \end{tabular}
    \caption{\justifying
    \textbf{\textit{Fiducial volume definition}}
    We apply the TPC fiducial volume cut for two charged tracks as defined in Ref.~\cite{T2K:2019jwa}, along with an additional cut specific to the HA-TPC used in this study. The beam propagates along the z-axis.}
    \label{tab:fid}
\end{table}

\end{appendix}
 
\end{document}